\documentclass[conference]{IEEEtran}
\IEEEoverridecommandlockouts
\usepackage{bbding}
\usepackage{cite}
\usepackage{amsmath,amssymb,amsfonts}
\usepackage{algorithmic}
\usepackage{graphicx}
\usepackage{textcomp}
\usepackage{hyperref}
\usepackage{booktabs, multirow} 
\usepackage{soul}
\usepackage[table]{xcolor} 
\usepackage{changepage,threeparttable} 

\newcommand{\larry}[1]{\textsf{\color{purple}{[{Larry: #1}]}}}

\def\BibTeX{{\rm B\kern-.05em{\sc i\kern-.025em b}\kern-.08em
    T\kern-.1667em\lower.7ex\hbox{E}\kern-.125emX}}
    
\begin{document}

\title{Expectations Versus Reality: Evaluating Intrusion Detection Systems in Practice}




\author{
\IEEEauthorblockN{Jake Hesford\IEEEauthorrefmark{1},
Daniel Cheng\IEEEauthorrefmark{1},
Alan Wan\IEEEauthorrefmark{1},
Larry Huynh\IEEEauthorrefmark{1},
Seungho Kim\IEEEauthorrefmark{2},
Hyoungshick Kim\IEEEauthorrefmark{2}, and
Jin B. Hong\IEEEauthorrefmark{1} \Envelope}
\\
\IEEEauthorblockA{\IEEEauthorrefmark{1}University of Western Australia, Australia\\ Emails: 21683564@student.uwa.edu.au, 23126543@student.uwa.edu.au, 23072152@student.uwa.edu.au,\\larry.huynh@uwa.edu.au, jin.hong@uwa.edu.au \Envelope} \\ \IEEEauthorblockA{\IEEEauthorrefmark{2}Sungkyunkwan University, Republic of Korea\\ Emails: kimsho98@naver.com, hyoung@skku.edu} 
}
\maketitle

\begin{abstract}
Our paper provides empirical comparisons between recent IDSs to provide an objective comparison between them to help users choose the most appropriate solution based on their requirements. Our results show that no one solution is the best, but is dependent on external variables such as the types of attacks, complexity, and network environment in the dataset. For example, BoT\_IoT and Stratosphere IoT datasets both capture IoT-related attacks, but the deep neural network performed the best when tested using the BoT\_IoT dataset while HELAD performed the best when tested using the Stratosphere IoT dataset. So although we found that a deep neural network solution had the highest average F1 scores on tested datasets, it is not always the best-performing one. We further discuss difficulties in using IDS from literature and project repositories, which complicated drawing definitive conclusions regarding IDS selection. 
\end{abstract}

\section{Introduction} 

There are many different intrusion detection systems (IDS) proposed over the years, advancing the state-of-the-art with high-performance measures reported \cite{khraisat2019survey, yang2022systematic}. However, it is also a challenge when trying to compare them and choose the best one for your needs, because there is no standardisation due to the complexity of the environment that these IDSs were designed for. In order to determine to what degree IDSs can be adapted to different environments, we compare their performance across common Network Intrusion Detection Systems (NIDS) datasets. This approach aims to provide a more standardized basis for comparison, taking into account different variables such as attack types, networking technology, and network environments.


There are several key challenges when standardising IDSs for comparison. The first one is using different datasets for testing. IDS solutions are often tested on diverse datasets, each with its unique characteristics and labelling methodologies. This variability complicates direct comparisons between different IDSs, as each system may be optimised for specific types of data \cite{azam2023comparative}. Furthermore, the prevalent use of non-standardised datasets in testing makes it difficult to assess the generalisability and robustness of these systems. Where IDSs were tested on datasets constructed by the IDS developers, the performance seen in these tests may not generalise well when testing against other datasets \cite{ahmad2022comprehensive}.

The next significant challenge arises from the number of configuration options available in IDSs. These options, while allowing for customisation, can also lead to a lack of clarity in the evaluation methodology. Without a standardised approach to configuring these systems, results can vary widely, making it hard to determine the true effectiveness of an IDS under different scenarios \cite{song2021analysis}.
One important factor in this particular area is that it remains unclear whether this is primarily due to the dataset, or the IDSs. Even for the most well-cited IDS datasets, there is literature available which questions the correctness of the data available. The intensive and complex process of dataset creation can often lead to error-prone output, presenting challenges in standardised testing if it does not also include extensive data wrangling and modification alongside it \cite{liu2022error}.

Another challenge is that IDSs commonly either take packets or flows. Where a dataset does not contain both of these formats, adapting it into the form expected by a given IDS is non-trivial, where the expected format is not the one provided by the dataset authors. This discrepancy presents challenges in obtaining satisfactory results when an IDS and dataset are incompatible without significant processing \cite{khraisat2019survey}. Our evaluation process was further complicated by the necessity of converting these datasets into formats compatible with various IDS solutions. This data wrangling could amplify the errors and inconsistencies inherent in the datasets. Such transformation processes can introduce additional noise and inaccuracies, thereby skewing the results of the IDS evaluations. For other important historical datasets that have been used to benchmark many IDSs (such as NSL-Cup 99), the pcap files may not be available at all. Where machine learning and autoencoder-based IDSs relied on particular features to make their assessment, there were difficulties where these files were unavailable or crucial features were not extracted, and utilising these systems became problematic \cite{khraisat2019survey}.

Furthermore, the effectiveness of autoencoder-based IDSs, which rely on temporal and benign data to function optimally, is significantly hindered when pcap files are not categorised by scenario or if specific benign traffic data is absent. These systems require a baseline of ‘normal' traffic against which to compare potentially malicious activity. Without this, training these models to accurately detect anomalies is challenging. The absence of scenario-specific or benign traffic data limits the ability of these systems to establish a normative profile of network activity, which is crucial for their anomaly detection mechanisms. In these cases, we were able to attempt to train the models on initial benign traffic in the dataset, but this often did not result in adequate performance and may not accurately represent a proper ‘baseline’ of traffic in the scenario \cite{kitsune}.



Considering these issues, we propose a pipeline for the effective comparison of IDSs. Machine-learning based network IDSs (NIDS) have been increasing in popularity due to their ability to see more complex patterns, and thus display higher protection against unknown attacks than traditional IDSs \cite{kitsune}. These benefits also come with an increase in complexity. In order to determine how flexible these systems are, we focus on NIDSs for the purpose of this study. To evaluate how well these NIDSs generalise, we use five NIDS datasets.

\section{Related Work}

When assessing NIDSs, there are two main components to be considered --- how an IDS performs with a given dataset, and how datasets perform when run through a given NIDS. Concerning the latter component, work has been done in critically analysing the features of popular IDS datasets. Binbusayyis and Vaiyapuri \cite{binbusayyis2019identifying} sought to find the optimal feature sets of 4 popular datasets using an ensemble method of statistical filters. Layeghy et al. \cite{layeghy4141050benchmarking} conducted an in-depth analysis of the statistical properties of benign traffic in the CICIDS2017, UNSW and TON-IOT datasets, and compared them with two real-world datasets. They found significant differences in the statistical features between the synthetic and real-world datasets and concluded that the evaluation of NIDS algorithms on synthetic datasets does not guarantee performance in real-world scenarios. Ghurab et al. \cite{benchmark_datasets} took an extensive look at the different NIDS datasets used for benchmarking, concluding that using recent datasets may be recommended due to their wider attack coverage, but note that all datasets may be appropriate in differing circumstances. However, since the datasets are not run through any IDSs in this study, there is no clear framework or pipeline provided that could be used to directly compare the performance of different IDSs on these datasets, which is crucial information for users who wish to adopt and use those IDSs in practice. Some other works have also looked at benchmarking IDSs, but many start at the evaluation steps with all common datasets and IDSs setup \cite{CardenasBS06, MilenkoskiVKAP15}. Furthermore, some works propose frameworks for this benchmarking, but do not practically implement it \cite{AyoubiBJST22}. Starting from this idealised view does not address the practical setup issues we encountered. Maseer et al. \cite{benchmark_cicids} investigated the performance of various machine learning algorithms, both supervised and unsupervised, on the CICIDS2017 dataset. This study evaluated the performance of various classical ML algorithms on the web-based attacks within CICIDS2017. Antunes et al. \cite{antunes2022benchmarking} evaluate the CSE-CIC-IDS2018 dataset and benchmark common deep learning methods, such as LSTM and CNN. However, these studies each consider a different single dataset in their benchmarking process, rather than a comprehensive evaluation across multiple datasets. As mentioned above, the characteristics of datasets differ, and it is suggested to use more recent datasets to observe results closer to the practical settings.

There were some work that focused on identifying the limitations of current IDS research. Ahmad et al. \cite{ahmad2022comprehensive} identified a prevalent trend in IDS research where proposed solutions, developed and evaluated using a single model and dataset, often exhibit inherent bias. The shortage of reliable, real-world datasets has been cited to be a contributor to this issue \cite{ahmad2022comprehensive, sarhan2021netflow}. This specialisation can thus result in significantly reduced performance when the model is applied to different network datasets. Our work extends these findings by critically evaluating actual research and open-source IDS implementations across multiple datasets, offering a more comprehensive assessment of their practical performance and accessibility.


\section{IDS Analysis Pipeline}
For analysing and comparing Network IDSs (NIDS), we first select recent IDSs from the literature and public repository. Due to the large volume of new IDSs, we limit the selection, which is further described in Section \ref{sec:ids_selection}. Next, we select datasets for testing the selected IDSs, which is shown in Section \ref{sec:dataset_selection}.

\subsection{IDS Selection}
\label{sec:ids_selection}
The first step was to select the NIDSs to be evaluated, and Table \ref{tab:ids_selection} provides an overview of the examined IDSs and our selection/exclusion of them based on the criteria. 
In order to complete this, various criteria were used. We dealt with two main IDS categories - academic and non-academic. This was done in order to determine if there was a noticeable variation in usability or performance between peer-reviewed (academic) studies and practically applicable public (non-academic) systems. With academic systems, one might expect they would be more on the experimental side, thus possibly showing a higher variance in results or better results at the cost of setup/unknown stability. Further, publicly developed NIDSs could feasibly sacrifice some performance optimisations in favour of more stable releases, or be older and more highly trusted to be used in industry. We determined these trade-offs could be worth comparing to academic systems.
We used differing criteria for the two types according to their respective properties. 

For academic NIDS, the criteria were as follows; 
\begin{enumerate} 
    \item \textbf{Recency:}
    Papers selected had to be published within the last 5 years to capture recent and relevant insights into a rapidly evolving field of research.
    \item \textbf{Code Availability:}
    Each paper had to have the IDS code attached and available, usually through a GitHub Repository. Repositories with unavailable code could not be used as we could not run the analysis process on them.
    \item \textbf{ML-Oriented:}
    The IDSs had to use a machine learning-based detection scheme. ML NIDSs are a rapidly growing and highly promising section of the field as discussed in the introduction.
    \item \textbf{Reliability:}
    Publisher reliability was used, evaluating the ranking of each studies associated journal or conference. Preference was given to papers with more reputable publishers, as reliable conferences typically yield higher quality NIDS comparisons.
    \item \textbf{Usability:} 
    The NIDSs were also assessed by the level of difficulty required to run the NIDS out of the box. A study was prioritised if the code could simply be cloned and run directly, producing similar results to those the academic paper cited. An easily usable study would be more likely to see wider use, and thus be more worthy of inclusion in this assessment of NIDS performance in broader contexts. Furthermore, a well produced, coherent and simple NIDS interface tends to indicate a higher quality of system and design. We made minimal changes such as updating deprecated library versions and changing absolute path locations, but an IDS was invalidated if it was unable to be run following these changes. 
        
\end{enumerate}

There were a large number of NIDSs that satisfied the first four criteria, but many fell short on the fifth (Usability). Due to absence of provided virtual environments or interpreter versions, many of the NIDSs could not be run and were invalidated. In the case that no environment was provided, we attempted to run the system and determine a compatible environment. However, this was often complicated by resulting package incompatibilities between versions, such as between Keras and Tensorflow, containing the necessary functionality.

Following this process, we selected the remaining academic NIDSs that satisfied all of the necessary criteria; Kitsune \cite{kitsune}, HELAD \cite{helad} and Deep Neural Networks (DNN) \cite{sdnn}. 

Kitsune is an online, unsupervised, plug-and-play NIDS leveraging an ensemble of autoencoders. This paper was selected due to its' popularity, with around one thousand citations, along with its' adherence to the other criteria. 

HELAD built on the works of the Kitsune IDS, and along with its adherence to the criteria, was selected to provide a comparison to the popular Kitsune IDS to determine its difference in performance. 

The DNN study \cite{sdnn} compared the performance of various classical machine learning algorithms, and also established that they found a deep neural network of 3 layers was the optimal dimensions for their study. Past meeting all the other criteria, the broad coverage of this study was the primary reason for its inclusion.

For non-academic NIDSs, the criteria were as follows; 
\begin{enumerate} 
    \item \textbf{Code Availability:} The code had to be publicly available, primarily through GitHub repositories.
    \item \textbf{Popularity:} The GitHub repositories had to have over 250 stars and 100 forks. More popular repositories would tend to be of a higher quality, or at least exhibit simpler interfaces and be adaptable for wider use cases.
    \item \textbf{Proper Documentation:} Accurate and detailed information in relation to error states, machine learning mechanisms, and architecture must be available. As these public projects did not have directly connected papers attached, sufficient documentation surrounding setup and usage instructions has to be provided in order to verify expected results and run the system.
    \item \textbf{Ongoing Support:} There must be evidence of continuous maintenance for the NIDS, determined by the presence of active contributions to its source code. Regularly updated repositories would tend to reflect newer advancements, or at least be operable on recent devices.
    \item \textbf{Usability:} Similar to the academic NIDS selection criteria, the code is required to run without significant alteration.
\end{enumerate}

We selected one IDS from this process - Stratosphere Linux IPS (Slips) \cite{slips}. Slips claimed to be a behavioural-based intrusion detection and prevention system that uses machine learning algorithms to detect malicious network traffic. Version 1.0.7 of Slips was used for testing and evaluation but new versions are constantly being released.

As previously mentioned, one of the key issues in IDS selection was being able to actually run the systems without significant issues. But due to the dependencies, one of the the primary challenges was determining the optimal versions of library functions and architectures used. Ideally, virtual environments are provided to simplify this process and inform users precisely what setup is optimal. However, these proved to be frequently absent within academic IDSs, which led to us being unable to reproduce the authors' systems. This meant that when trying to run them, we encountered various errors as shown in Table 1.

\begin{table*}
\caption{IDSs investigated - green systems were used in the study, with red systems excluded for the displayed reason.} 
\label{tab:ids_selection}
\centering
\small
\begin{tabular}{|l|r|p{3.3cm}|p{3.8cm}|p{4cm}|}
\toprule
NIDS & Year & Dataset & Source & Usability/Issues \\
\midrule
\hline
\cellcolor[HTML]{b6d7a8}Deep Neural Network (DNN) \cite{sdnn} & 2018 & KDDCup-`99’ & Conference: ICCCNT & Used in Paper \\
\hline
\cellcolor[HTML]{b6d7a8}Kitsune \cite{kitsune} & 2018 & Custom IoT Dataset & Conference: NDSS & Used in Paper \\
\hline
\cellcolor[HTML]{a9d08e}HELAD \cite{helad}& 2020 & CICIDS2017 & Journal: MDPI Informatics & Used in Paper \\
\hline
\cellcolor[HTML]{f4cccc}Multiclass Classification \cite{multiclass_class} & 2020 & ASNM Datasets & Conference: DSAA & Vague dependencies in provided repository,  
"ValueError on converting string to complex in ASNM-TUN.py"\\

\hline
\cellcolor[HTML]{f4cccc}ARTEMIS.\cite{mqtt_deep} & 2021 & Custom Dataset & Conference: LATINCOM & Code error \\
\hline
\cellcolor[HTML]{f4cccc}Dense-Attention-LSTM, DAL \cite{cao2021network} & 2021 & UNSW-NB15 & Conference: IWCMC & Dependency errors \\
\hline
\cellcolor[HTML]{f4cccc}I-SiamIDS \cite{bedi2021siamids} & 2021 & CICIDS, NSL-KDD & Journal: Applied Intelligence & Type error \\
\hline
\cellcolor[HTML]{f4cccc}SecureTea \cite{securetea} & 2021 & N/A & GitHub & Dependency errors \\
\hline
\cellcolor[HTML]{f4cccc}AutoML \cite{yang2022iot} & 2022 & CICIDS2017, IoTID20 & Journal: Engineering Applications of Artificial Intelligence & IDS code not provided \\
\hline
\cellcolor[HTML]{f4cccc}Deep Belief Networks NIDS \cite{belarbi2022intrusion} & 2022 & CICIDS2017 & Conference: SciSec & Invalidated by dependency errors in provided repository: "Tensors found on two or more devices"\\
\hline
\cellcolor[HTML]{f4cccc}RIDS \cite{saini2022rids} & 2022 & Custom Dataset & Conference: GLOBECOM & Provided Out of memory\\
\hline
\cellcolor[HTML]{a9d08e}StratosphereIPS (Slip) \cite{slips} & 2022 & N/A & GitHub & Used in Paper \\
\hline
\cellcolor[HTML]{f4cccc}IDS-ML \cite{yang2022ids} & 2022 & CICIDS2017 & Journal: Software Impacts & Runtime errors \\
\hline
\cellcolor[HTML]{f4cccc}xNIDS \cite{wei2023xnids} & 2023 & Mirai, CICDoS2017, NSL-KDD & Conference: USENIX Security & Did not propose a directly usable NIDS, so was not appropriate.\\
\hline
\cellcolor[HTML]{f4cccc}Suricata \cite{suricata} & 2023 & N/A & GitHub & Unable to verify any use of ML \\
\bottomrule
\end{tabular}
\end{table*}

\subsection{Dataset Selection}
\label{sec:dataset_selection}
In our study, the selection and evaluation of datasets played a crucial role in assessing IDSs. We attempted to focus on datasets that varied in terms of attack types, traffic origin, protocols, and other key factors to ensure a comprehensive analysis.

\subsubsection{Selection Criteria}

In order to provide a simple and structured initial evaluation of the datasets, we were guided by the selection methodologies outlined by previous works \cite{sharafaldin2017towards} ensuring a robust and systematic approach. Additionally, we have included "Popularity" as a criteria, focusing on well-cited datasets to ensure that we are working with data that has been widely recognised by the research community. The criteria are as follows: 

\begin{enumerate}
    \item \textbf{Representation of Modern Network Threats:} We prioritised datasets that included current and emerging attack types, as well as those that represented modern network traffic patterns, to better reflect the current landscape of potential threats. As attacks evolve, IDSs can become outdated if they are not updated to reflect modern threats. This is crucial for developing IDS software that can effectively mitigate real-world security threats, though may be mitigated by ML and AI systems.
    \item \textbf{Realism and Diversity:} Datasets that closely mimic real-world network environments and offer a diverse range of traffic types and attack scenarios were favoured. A broader range of attack types representing modern traffic will provide a better look at the adaptability of the systems to a range of traffic types.
    \item \textbf{Availability and Quality of Data:} The datasets needed to be either publicly accessible, or accessible through permission granted by the dataset authors, and of high quality, with minimal errors or inconsistencies. If the datasets were not available they could not be analysed, and thus would not have been viable for the study.
    
    \item \textbf{Popularity:} We prioritised datasets that are commonly used within the research community for evaluating methodologies. More popular datasets would not only indicate higher levels of use and thus quality, but be more likely to contain well-formatted and consistent data that would be more easily transferable.
\end{enumerate}

By applying these criteria, we were able to select datasets that not only provided a comprehensive and realistic environment for testing but also aligned well with the capabilities and requirements of the IDSs being evaluated. The selection of datasets evolved slightly over the experimentation period as certain selected datasets became difficult to process and use on the selected IDSs.

\subsubsection{Evaluated Datasets Used in Results}

Datasets like CICIDS2017 and UNSW-NB15 provide a balanced and realistic representation of network traffic and modern attack types. Their labelled nature and comprehensive feature sets made them more suitable for our analysis. The IoT-specific datasets (Stratosphere IoT, Mirai, BoT-IoT, and ToN-IoT) were chosen for their relevance to current and emerging threats in the IoT domain, an area of growing importance in network security and one where many new ML IDSs are focused. Table \ref{tab:datasets} provides an overview of the datasets that were used in our evaluations. 

\begin{table}
\centering
\caption{Datasets Used for Evaluation}
\label{tab:datasets}
\begin{tabular}{|p{1.4cm}|p{3cm}|p{3cm}|}
\hline
\textbf{Dataset} & \textbf{Characteristics} & \textbf{Relevance and Reason for Selection} \\ \hline
CICIDS2017 \cite{sharafaldin2018toward} & Includes traffic from various devices and operating systems. Labelled with 80 features over 5 days. & Comprehensive range of attacks; ideal for evaluating modern IDSs due to diversity and extensive feature set. \\ \hline
UNSW-NB15 \cite{moustafa2015unsw} & Generated by ACCS with 49 features and 9 attack types over 2 days. & Represents a wide spectrum of contemporary attack types, providing a broad base for IDS effectiveness testing. \\ \hline
Stratosphere IoT CTU \cite{iotctu} & Focuses on IoT network traffic, with realistic threat and behaviour representation. & Essential for understanding IDS effectiveness in IoT environments due to its focus on realistic IoT-specific threats. \\ \hline
Mirai (Kitsune) \cite{kitsune} & Data specific to Mirai botnet attacks, used with the Kitsune IDS. & Demonstrates significant Mirai threat in IoT, allowing for practical assessment of IDS capabilities against IoT botnets. \\ \hline
BoT-IoT \cite{ashraf2021iotbot} \& ToN-IoT \cite{moustafa2021new} & Encompasses legitimate and emulated IoT network traffic. & Offers a balanced view of IDS performance in IoT settings, serving as a robust alternative to the Kitsune dataset. \\ \hline
\end{tabular}
\end{table}

\subsubsection{Examined Datasets Not Included in Experimentation}

In addition to selecting datasets that were most suitable for our study, we also considered several datasets that we ultimately did not use. The decision to exclude certain datasets was based on a set of criteria developed to ensure the most effective and comprehensive evaluation of the IDS solutions. Below is an overview of the datasets we did not use, the reasons for their exclusion, and the criteria we employed in selecting datasets. Table \ref{tab:excluded_datasets} provides an overview of the datasets that were excluded in our evaluations. 

\begin{table}
\centering
\caption{Datasets Considered but Not Used for Evaluation}
\label{tab:excluded_datasets}
\begin{tabular}{|p{1.4cm}|p{3cm}|p{3cm}|}
\hline
\textbf{Dataset} & \textbf{Characteristics} & \textbf{Relevance and Reason for Exclusion} \\ \hline
KDD-Cup \cite{kddcup} \& NSL-KDD \cite{tavallaee2009detailed} & Historically significant but outdated, lacking pcap files. & Not representative of current network behaviours; incompatible with selected IDSs due to lack of pcap files. \\ \hline
CAIDA \cite{passive_merged_pcap} & Limited attack diversity and lacks full network data, unlabelled. & Unable to train auto-encoders on the dataset due to lack of labelled results. \\ \hline
CIDDS \cite{ring2017flow} \cite{ring2017creation} & Designed for anomaly-based network security. & Not widely used in literature, suggesting potential limitations for analysis. \\ \hline
ISCX2012 \cite{shiravi2012toward} & Older dataset without features & Due to lack of features, other datasets were determined to be more suitable  \\ \hline
CICIDS2019 \cite{sharafaldin2019developing} & Modern DDoS Dataset containing a variety of DDoS attack types. & Strong modern DDoS dataset, but was not chosen due to the specific nature of attacks when compared to more general datasets used. \\ \hline
Kyoto \cite{song2011statistical} & Realistic, unsimulated dataset derived from diverse honeypots. & Offers a different perspective to generated datasets, but not highly cited.  \\ \hline
LBNL \cite{pang2005first} & Heavy anonymisation and absence of payload data. & Limits the depth of analysis for IDSs, making it less favourable for in-depth IDS evaluation. \\ \hline
CICIDS2018 \cite{sharafaldin2018toward} & Diverse traffic and heavy volume without specific pcaps. & Only available as 250gb file, data wrangling complexity and volume make processing unwieldy. \\ \hline
ASNM Datasets \cite{homoliak2020asnm} & NIDS anomaly-based datasets developed for machine learning. & Attack diversity is limited and not as well-cited as many other options. \\ \hline
IoTID \cite{ullah2020scheme} & Newer IoT Dataset that aimed to target new IoT intrusion methods & Narrow dataset that is not as popular as the other chosen IoT datasets. \\ \hline
CICDOS2017 \cite{jazi2017detecting} & DoS Dataset generated by CIC based on the ISCX dataset & Narrow dataset without attack diversity of CIC dataset from the same year. \\ \hline
\end{tabular}
\end{table}


As our experimentation began, we also had to begin to consider the availability and adaptability of the data to work with the IDSs under test. We faced challenges with datasets that were not standardised, lacked certain features, or were formatted in ways incompatible with our IDS solutions. For example, datasets like KDD’99 and NSL-KDD, while historically significant, posed difficulties due to their outdated nature and lack of pcap files. Similarly, datasets like CAIDA, while useful for DDoS attack analysis, offered limited diversity in attack types and were not labeled.

\section{Testing and Evaluation}

Our testing and evaluation methodology for IDSs was designed to provide a comprehensive assessment of each system's effectiveness. This section outlines the approach taken to test each IDS against the selected datasets, the metrics and criteria used for performance evaluation, and the approach for handling and interpreting the results.

\subsection{Methodology for Testing Each IDS}
\begin{enumerate}
    \item \textbf{Data Preprocessing and Sampling:} Significant preprocessing of the dataset files was necessary, including converting datasets into the required format and extracting relevant features for compatibility with each IDS. Random flow sampling was performed on these processed files when the size of dataset files inhibited complete testing. 
    \item \textbf{Handling Temporal Statistics:} After random sampling of the packets, the results were sorted by their timestamp. This ensured that the IDSs received data that preserved the temporal statistics of the input packets. This step was performed as the IDSs utilise the temporal statistics and trends of input packets.
    \item \textbf{IDS Configuration and Deployment:}  Each IDS was configured according to the standard setup instructions provided by the developers, without incorporating any customisations or optimisations that could enhance (or detract from) performance. The testing process involved processing the prepared datasets with these baseline configurations. The IDSs utilise machine learning algorithms to classify anomalies, and adjustments were not made to the model parameters shipped with the initial setup instructions, ensuring a uniform, out-of-the-box evaluation framework for all tested IDS solutions.
    \item \textbf{Anomaly Detection Threshold:} The anomaly detection threshold for each IDS was determined through a standardised process to ensure fairness in evaluation. This process involved identifying the threshold value that maximised the detection rate of anomalous packets while maintaining a tolerable level of false positives for the given results. The specific threshold value might differ across IDSs due to their varying sensitivity and detection mechanisms, but the methodology for determining this value remained consistent.
\end{enumerate}

\subsection{Evaluation Metrics and Criteria}
The performance of the IDS solutions was evaluated using the standard evaluation metrics: accuracy, precision, recall, and F1 scores.

\section{Results}
The overall results of empirical analysis are shown in Table \ref{tab:performance}. The source code used for our evaluation is available from \textit{hidden}.

\begin{table}[!h]
\caption{Performance Results for Tested IDSs and Datasets}
\centering
\small
\renewcommand{\arraystretch}{1.3}
\begin{tabular}{|l|c|c|c|c|c|c|}
\hline
\textbf{Dataset} & \textbf{Acc.} & \textbf{Prec.} & \textbf{Rec.} & \textbf{F1}\\
\hline
\multicolumn{5}{|l|}{IDS: Kitsune} \\
\hline
UNSW-NB15       & 0.6954 & 0.0221 & 0.2136 & 0.0401 \\
BoT\_IoT        & \textbf{0.9923} & 0.8153 & 0.8609 & 0.8375\\
CICIDS2017      & 0.5540 & 0.0109 & 0.9753 & 0.0216 \\
Stratosphere    & 0.9921 & 0.9981 & 0.9027 & 0.9480\\ 
Mirai           & 0.8902 & \textbf{0.9999} & 0.8788 & 0.9354\\ 
\hline
Average:        & 0.8248 & 0.5693 & 0.7663 & 0.5565 \\
\hline
\multicolumn{5}{|l|}{IDS: HELAD} \\
\hline
UNSW-NB15       & 0.9717 & 0.0201 & 0.0107 & 0.0140 \\
BoT\_IoT        & 0.9793 & 0.6916 & 0.9011 & 0.7826 \\
CICIDS2017      & 0.6437 & 0.9682 & 0.3706 & 0.5360 \\
Stratosphere    & 0.9846 & 0.9805 & \textbf{1.0000} & {\color{blue}0.9902} \\
Mirai           & 0.8898 & 0.9939 & 0.8786 & 0.9327 \\
\hline
Average:        & 0.8938 & 0.7284 & 0.6322 & 0.6511 \\
\hline
\multicolumn{5}{|l|}{IDS: DNN} \\
\hline
UNSW-NB15       & 0.9820 & 0.9820 & \textbf{1.0000} & {\color{blue}\textbf{0.9910}} \\
BoT\_IoT        & 0.9770 & 0.9770 & \textbf{1.0000} & {\color{blue}0.9884} \\
CICIDS2017      & 0.9800 & 0.9800 & \textbf{1.0000} & {\color{blue}0.9899} \\
Stratosphere    & 0.2110 & 0.2110 & \textbf{1.0000} & 0.3485 \\
Mirai           & 0.9060 & 0.9060 & \textbf{1.0000} & {\color{blue}0.9507}\\
\hline

Average:        & 0.8112 & 0.8112 & 1.0000 & 0.8537 \\
\hline
\hline
\multicolumn{5}{|l|}{IDS: Slip} \\
\hline
UNSW-NB15       & 0.8735 & 0.0000 & 0.0000 & 0.0000  \\
BoT\_IoT        & 0.0018 & 0.0000 & 0.0000 & 0.0000  \\
CICIDS2017      & 0.9370 & 0.0037 & 0.0447 & 0.0068 \\
Stratosphere    & 0.6745 & 0.8809 & 0.4739 & 0.6163 \\
Mirai           & 0.8040 & 0.1243 & 0.0159 & 0.0282 \\
\hline
Average:        & 0.6582 & 0.2018 & 0.1069 & 0.1303  \\
\hline
\end{tabular}
    \begin{tablenotes}
      \footnotesize
      \item *\textbf{Bolded value}: the highest value of all IDSs for the metric column.
      \item **{\color{blue}Font colour blue}: the highest F1 score of all IDSs for the dataset.
    \end{tablenotes}
\label{tab:performance}
\end{table}

These results shows that using the DNN \cite{sdnn} has the most versatility across all datasets tested, achieving the highest average F1 score amongst all tested IDSs. However, it did perform worse on the Stratosphere dataset, indicating that it may not be the most optimal solution depending on user requirements. Given the Stratosphere dataset focuses on IoT network traffic, at the same time the BoT-IoT also focuses on IoT network traffic, the performance difference between the two datasets requires further investigation in terms of dataset differences to better understand why we observe such differences in the performances. However, this is outside the scope of this paper. 
On the other hand, HELAD achieved the highest average Accuracy, but given the datasets are not fully balanced, this is not the best indicator of which IDS performs the best. Hence, when proposing IDSs, the authors should present various metrics to measure the performance of their IDSs for ease of comparison against others.

Several factors identified throughout the testing process can explain the sub-optimal results observed in some cases: 
\begin{enumerate}
    \item \textbf{Inconsistent Performance Across Datasets:} The results show significant variation in the performance of different IDS models across datasets. For instance, Kitsune demonstrated high accuracy in the IoT-focused datasets BoT-IoT and Stratosphere (0.9923 and 0.9921 respectively), but significantly lower accuracy in CICIDS2017 (0.5540). This disparity suggests that the effectiveness of an IDS model can be highly dependent on the specific characteristics of the dataset.
    \item \textbf{Overfitting to Specific Dataset Characteristics:} Kitsune's high accuracy in IoT datasets (BoT-IoT and Stratosphere) versus its lower performance in CICIDS2017 may also suggest overfitting to certain dataset traits, impairing its effectiveness in more diverse network conditions.


    \item \textbf{High False Positives/Negatives in Certain Scenarios:} The performance metrics, particularly the precision and recall values, indicate potential issues with false positives and negatives. For example, in the CICIDS2017 dataset, HELAD showed a high precision of 0.9682, but a much lower recall of 0.3706, implying a tendency to miss actual attacks (high false negatives).
    \item \textbf{Dataset and IDS Model Compatibility:} The specific structure of individual datasets could have impacted the IDS models differently. The DNN showed exceptional performance on UNSW-NB15, BoT-IoT, and CICIDS2017 (with F1 scores above 0.98) but struggled significantly with the Stratosphere dataset (F1 score of 0.3485). This suggests that specific features of Stratosphere may not be well-handled by this particular model and may warrant further exploration.
    \item \textbf{Impact of Preprocessing on Model Efficacy:} The preprocessing steps necessary for compatibility could have differentially impacted the models. For instance, the poor performance of the DNN on Stratosphere (0.2110 accuracy) compared to its high performance on other datasets may also indicate preprocessing issues specific to this dataset.
    \item \textbf{Lack of Representative Benign Traffic:} Some datasets do not explicitly provide a benign traffic baseline. High precision but low recall in models like HELAD on datasets such as CICIDS2017 may indicate insufficient representation of benign traffic. This can lead to higher false positives, as the model is unable to recognise normal network behaviours effectively.
\end{enumerate}

Through this comprehensive analysis, we aimed to provide not just an empirical comparison of different IDS solutions but also a deeper understanding of the factors influencing their performance, essential for advancing network security and guiding the development of more effective IDS.

\section{Discussion}
In our evaluation, we encountered a series of insights and challenges pivotal for understanding the current state and future direction of IDS technologies. This discussion section delves into the key findings, their implications, and the broader context of network security.

\subsection{Interpretation of Results}

Our results indicated a significant variation in the performance of different IDS solutions across various datasets. Notably:

\begin{enumerate}
    \item \textbf{Performance Variability:} The effectiveness of IDS varied significantly depending on the dataset used. This variability underscores the importance of selecting diverse datasets for IDS evaluation, ensuring that a diverse set of attack scenarios are considered and potentially different areas of strength based on both the IDS and its configuration options.
    \item \textbf{Dataset-Specific Challenges:} The performance of IDS solutions may be adversely affected by datasets that lack a representative benign traffic sample. The evaluation of the system may also be affected by the variety of attack types present in the dataset. This highlights the need for diverse and comprehensive datasets in IDS testing.
    \item \textbf{Preprocessing Impact:} The preprocessing steps necessary for formatting and compatibility may play a crucial role in IDS performance. In some cases, this could lead to data loss or the introduction of errors, impacting the accuracy of the systems.
    \item \textbf{Challenges in Handling Real-World Network Dynamics:} The struggle of some models with datasets that include a wider range of traffic patterns highlights the difficulty in adapting to the complexities of real-world network dynamics, especially without custom configuration of IDS solutions.
    \item \textbf{Adaptability to Evolving Threats:} The results also highlight the challenge for IDS solutions to adapt to broad and evolving threats. Static datasets may not encompass emerging attack vectors, necessitating continual updates and testing against new threat scenarios. The addition of UNSW and CICIDS2017 in our results highlighted potential gaps in the capabilities of IDSs, but without these, we may have presented inflated results.
    \item \textbf{Need for More Comprehensive Testing:} The varied results across different datasets indicate the need for more comprehensive testing conditions to evaluate the robustness of IDS models
\end{enumerate}

\subsection{Insights from IDS Performance on Specific Datasets}

\begin{enumerate}
    \item \textbf{Dataset Compatibility:} The IDS solutions demonstrated high effectiveness when tested on datasets they were either designed for or shipped with. This suggests that these systems can be highly effective when the data characteristics align closely with their underlying detection algorithms.
    \item \textbf{Role of Benign Traffic Profile:} The strong results on the Stratosphere CTU IoT Dataset underscore the importance of having a well-defined benign traffic profile. This dataset, containing a realistic representation of IoT network traffic, including both normal and malicious activities, provided an ideal environment for IDS to distinguish between benign and malicious behaviours accurately.
    \item \textbf{Model Overfitting and Generalisation:} The results suggest a potential issue of overfitting in some IDS models. For instance, a model showing high accuracy in specific datasets (like Kitsune in BOT IoT) but performing poorly in others (like CICIDS2017) may indicate a lack of generalisation capability.
    \item \textbf{False Positives/Negatives Concerns:} Disparities in precision and recall metrics across datasets indicate issues with false positives or negatives. For example, HELAD's high precision but low recall in CICIDS2017 points to a tendency to miss attacks, emphasising the need for a balanced approach to anomaly detection.
    \item \textbf{Customisation and Tuning:} These observations suggest that IDS solutions, when appropriately customised and tuned to specific network environments and traffic profiles, can offer robust detection capabilities. This customisation is crucial to the eventual performance of the system, though. Plug-and-play IDSs may not offer adequate performance in certain circumstances.
\end{enumerate}

\subsection{Practical Implications for Deployment}

Our findings have several practical implications for the deployment of IDS solutions in real-world scenarios:

\begin{enumerate}
    \item \textbf{Customisation for Network Environment:} The variability in performance across different datasets, such as the high F1 score of 0.9480 for Stratosphere and the low F1 score of 0.0216 for CICIDS2017 in the Kitsune IDS, underscores the need for tailoring or optimising of IDS solutions for specific network environments.
    \item \textbf{Need for Dynamic Testing Datasets:} The disparities in IDS performance across datasets, such as HELAD's high performance in a narrower dataset such as Stratosphere (F1 score of 0.9902) versus its lower performance in the more general UNSW-NB15 (F1 score of 0.0140), support calls for developing dynamic testing datasets that accurately simulate real-world conditions, including a mix of known threats and benign activities \cite{layeghy4141050benchmarking}.
    \item \textbf{Ongoing Adaptation and Update:} The landscape of network threats is continually evolving, both datasets and IDSs can quickly fall behind current trends, this necessitates regular updates and adaptations of both IDS solutions and testing datasets to new threats and network conditions \cite{khraisat2019survey}.
    \item \textbf{Holistic Security Approach:} The varying effectiveness of different IDS solutions (e.g., Slip's low average F1 score of 0.1303) suggests that relying solely on IDS is insufficient. Instead, IDS should be integrated into a multi-layered security approach for robust protection.
\end{enumerate}

\subsection{Future Directions and Recommendations}

Based on our study, we recommend the following directions for future research and development in IDS:

\begin{enumerate}
    \item \textbf{Code availability and documentation:} One critical issue was not being able to access all codes necessary to run an IDS, or the lack of documentation to perform trouble shooting or customisation of the IDS. As shown in Table \ref{tab:ids_selection}, many IDSs had error-based issues that rejected them from being analysed. We attempted to contact authors for all these occasions without success, hindering the use of these IDSs albeit their great performance reported in the paper.
    \item \textbf{Development of comprehensive datasets:} There is a growing need for more comprehensive and diverse datasets that accurately reflect current network environments and attack vectors, as many recent IDSs are based on machine learning techniques. As shown in Table \ref{tab:performance}, the IDSs performed differently even though the datasets were capturing the same type of attacks (e.g., Stratosphere and BoT-IoT). Moreover, the standardisation of datasets and IDS input data is also of paramount importance, as not all dataset creators provide the same information as others (e.g., pcaps only, flows only, unlabelled pcaps, etc.). 
    \item \textbf{Virtualisation:} The analysis is difficult to perform due to different dependency requirements for IDSs. By providing a virtual environment, we can significantly reduce the code-based errors to perform evaluation. However, we have not seen any IDS solutions provided in a virtualised environment, hindering the evaluation processes. 
\end{enumerate}

\section{Conclusion}
Creating a versatile IDS is a challenging task due to the complex nature of intrusions and the environment the IDS operates. Our IDS analysis process provides an overview of how different recent IDSs performed when considering various datasets. Amongst them, the DNN solution \cite{sdnn} was the most versatile IDS, achieving the highest average F1 score of 0.8537 (followed by HELAD with F1 score of 0.6511). However, HELAD was better performing considering the test dataset Stratosphere, indicating that depending on the use cases, the best IDS to choose will differ. Further, our results and experience showed a gap between research and practical applications, with many challenges faced when trying to use other IDSs from both academia and public projects, as well as the discrepancies in datasets for evaluating IDSs. Therefore, we also suggest future IDS developers various aspects for consideration in order to better standardise the evaluation and adoption of IDSs in the future, which will enable better utilisation of them by users who can better understand the capabilities and limitations of IDSs.





\newpage

\bibliographystyle{IEEEtran}
\bibliography{main.bib}

\end{document}